\documentclass{PoS}

\usepackage{amssymb,amsmath,amsthm}
\usepackage{feynmp}

\title{$K \to \pi \pi$ Amplitudes at Unphysical Kinematics Using Domain Wall
Fermions}

\ShortTitle{$K \to \pi \pi$ Amplitudes at Unphysical Kinematics Using Domain Wall
Fermions}

\author{\speaker{Matthew Lightman}\\
        Department of Physics, Columbia University, New York, NY 10027, USA\\
        E-mail: \email{lightman@phys.columbia.edu}}

\author{RBC and UKQCD collaborations}

\abstract{The use of chiral perturbation theory in extracting physical $K\to\pi\pi$ matrix elements from matrix elements calculated at unphysical kinematics is outlined.  In particular, the possibility of utilizing pions with non-zero momentum in the final state, and of using partial quenching is discussed.  Preliminary (not physically normalized) $\Delta I=3/2$ (27,1) $K\to\pi\pi$ matrix elements are calculated  on the RBC/UKQCD $24^3 \times 64$, $L_s=16$ lattices, using 2+1 dynamical flavors and domain wall fermions, with an inverse lattice spacing of $a^{-1}=1.729(28)\text{ GeV}$.  Effective mass plots are presented for a light sea quark mass of $m_l^{\text{sea}}=0.005$, and various valence quark masses.  The plateaux are fit and $E_{\pi\pi}-m_K$ is extracted.}

\FullConference{The XXV International Symposium on Lattice Field Theory\\
                 July 30 - August 4 2007\\
                 Regensburg, Germany}

\begin{document}

\section{Introduction}

The decay $K \to \pi \pi$ is important to simulate on the lattice because QCD effects come into play since the typical energies are smaller than the scale $\Lambda_{QCD}$.  In particular, the direct CP violating parameter $Re\left(\epsilon'/\epsilon\right)$ can be found from calculations of $K \to \pi \pi$ matrix elements \protect\cite{CPPACS,RBC}.  In order to obtain reasonable precision in these calculations we plan to use 2+1 flavors of domain wall fermions (DWF) on a $24^3 \times 64$, $L_s=16$ lattice.

Accurate simulation of physical pions would require a larger box size than is currently available, therefore we make use of the tools of chiral perturbation theory ($\chi$PT).  $\chi$PT allows one to make predictions of the dependence of matrix elements on quark masses when they are close to the chiral limit.  It can even be applied to the case of unphysical kinematics.  Thus the best strategy is to simulate $K \to \pi \pi$ at unphysical kinematics, fit the results to $\chi$PT, and then use $\chi$PT to make a prediction for the physical values of the matrix elements.

\section{Four Quark Operators and $\chi$PT}

The weak interactions are included in the lattice QCD simulation by evaluating matrix elements of the effective Hamiltonian \protect\cite{Ciuchini,Buchalla}
\begin{equation}
\mathcal{H}_{\Delta S = 1}=\frac{G_F}{\sqrt{2}}\sum\limits_i V^i_{CKM}c_i(\mu)Q_i
\end{equation}
where $c_i(\mu)$ are Wilson coefficients and $\{Q_i,i=1,...,10\}$ are four quark operators.  Therefore we are interested in calculating matrix elements of the four quark operators $Q_i$ between a $K$ and a $\pi\pi$ state.  These operators can be split into $\Delta I=3/2$ and $\Delta I=1/2$ parts, called $Q_i^{3/2}$ and $Q_i^{1/2}$ respectively, where $\Delta I$ is the change in isospin induced by the operator.  They can then be further classified by how they transform under the chiral $SU(3)_L\times SU(3)_R$ symmetry, and the transformation properties (27,1), (8,8), and (8,1) are all found among various of the operators \protect\cite{CPPACS,RBC}.

In $\chi PT$ an effective Lagrangian \protect\cite{Georgi} is written in terms of the field
\begin{equation}
\Sigma=\exp\left[\frac{2i\phi^a\lambda^a}{f}\right]
\end{equation}
where the $\phi^a$ are the real pseudo-scalar meson fields.  The leading order part of the effective Lagrangian is
\begin{equation}
\mathcal{L}_{LO}=\frac{f^2}{8}Tr[\partial_\mu\Sigma\partial^\mu\Sigma]+\frac{f^2B_0}{4}Tr[\chi^\dag\Sigma+\Sigma^\dag\chi]
\end{equation}
where $\chi=diag(m_u,m_d,m_s)$ and $B_0=\frac{m_{\pi^+}^2}{m_u+m_d}=\frac{m_{K^+}^2}{m_u+m_s}=\frac{m_{K^0}^2}{m_d+m_s}$.  To represent the four quark operators in $\chi PT$, one looks at the operators that have definite transformation properties under isospin and $SU(3)_L\times SU(3)_R$, and forms operators out of the $\Sigma$ field that transform in the same way.  In general it will be possible to form more than one such operator, and thus a linear combination of all of these operators in which each operator is multiplied by an arbitrary coefficient is taken to represent the four quark operator \protect\cite{Bernard}.  These arbitrary coefficients are called low energy constants (LECs).  A $\chi PT$ expression for a $K \to \pi\pi$ matrix element of a four quark operator at next to leading order (NLO) and with pions in the final state having zero three-momentum, generally contains polynomials in the meson masses with order $m^2$ and order $m^4$ terms, as well as non-analytic terms in the meson masses known as chiral logarithms.  The coefficients of these terms contain LECs and other parameters in the Lagrangian.

\section{Extraction of the LECs}

Matrix elements of four quark operators can be computed on the lattice for several different quark masses.  The results can be fit to the $\chi PT$ formulae and the fit will yield LECs.  The strategy is to compute these matrix elements at unphysical kinematics in order to extract the LECs that appear in the $\chi PT$ expression for the physical matrix elements.  However, in order to be able to determine the necessary LECs uniquely with a limited number of gauge field ensembles one must either consider pions with non-zero momenta, or one must resort to partial quenching in which the masses of the quarks in the fermion determinant (sea quarks) are different from the masses of the propagating quarks (valence quarks).

\subsection{Pions with Non-Zero Momenta}

$\chi PT$ formulae for $K \to \pi\pi$ matrix elements with pions having non-zero momentum have been worked out by Lin et. al. \protect\cite{Lin1} and Laiho and Soni \protect\cite{Soni_Laiho2}.  One drawback of this method is that data with non-zero momentum tends to be very noisy, especially as the momentum increases.  There exist methods for dealing with this such as antiperiodic, and in general twisted boundary conditions.

\subsection{Partial Quenching}

The largest $24^3 \times 64$, $L_s=16$ RBC/UKQCD 2+1 flavour dynamical lattice ensembles currently available have sea quark masses $m_u^{\text{sea}}=m_d^{\text{sea}}=\{0.005,0.01\}$ and $m_s^{\text{sea}}=0.04$.  In general it is necessary to vary both the light quark and strange quark valence masses in order to extract LECs.  Laiho and Soni \protect\cite{Soni_Laiho1} have treated the case of partially quenched $\chi PT$ at NLO with sea quarks of equal mass.  However, partially quenched $\chi PT$ formulae at NLO for the case of unequal sea quark masses do not yet exist in the literature and are in the process of being calculated by Christopher Aubin, Shu Li, and Jack Laiho.  The current plan is to focus primarily on the partial quenching technique.  Non-zero momenta can then be incorporated as a consistency check of the LECs obtained from the former method.

Table 1 
gives a list of sets of quark masses that would be sufficient to extract the necessary LECs from an analysis of \protect\cite{Soni_Laiho1}.  However, it is not guaranteed that these masses, with the sea strange quark mass changed to $m_s^{\text{sea}}=0.04$ in all cases, will still be sufficient when one considers $\chi PT$ with unequal sea quark masses.  This analysis was done merely as an exercise to get a feeling for the number of combinations of masses that would be needed.
\begin{table}
\label{tb:Masses}
\caption{A list of sets of quark masses at which to evaluate several matrix elements, from which the LECs needed to calculate the physical $K\to\pi\pi$ matrix elements can be extracted.}
\begin{center}
\begin{tabular}{|c|c|c|c|c|}
\hline
$m_l^{\text{sea}}=m_s^{\text{sea}}$&$m_s^{\text{val}}$&$m_l^{\text{val}}$\\
\hline
0.01&0.01&0.01 \\
0.02&0.02&0.02 \\
0.02&0.01&0.01 \\
0.005&0.035&0.005 \\
0.01&0.02&0.01 \\
0.02&0.03&0.01 \\
\hline
\end{tabular}
\end{center}
\end{table}

\section{Matrix Element Calculations}

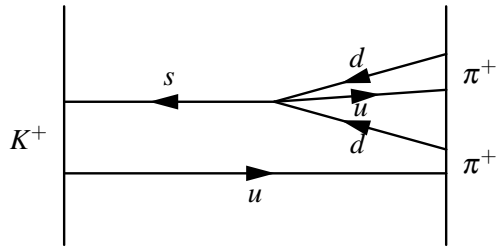
\begin{figure}[htp]
\centering
\setlength{\unitlength}{0.025in}
\begin{fmffile}{KPiPi}
\begin{fmfgraph*}(80,50)
\fmfkeep{KPiPi}
\fmfstraight
\fmfleftn{i}{4}
\fmfrightn{j}{6}
\fmf{plain,tension=1}{i1,i2}
\fmf{plain,label=$K^+$,label.side=left}{i2,i3}
\fmf{plain,tension=1}{i3,i4}
\fmf{plain,tension=1}{j1,j2}
\fmf{plain,label=$\pi^+$}{j2,j3}
\fmf{plain,tension=1}{j3,j4}
\fmf{plain,label=$\pi^+$}{j4,j5}
\fmf{plain,tension=1}{j5,j6}
\fmf{fermion,label=$s$,tension=3}{v1,i3}
\fmf{fermion,label=$d$,label.dist=3,label.side=left}{j3,v1}
\fmf{fermion,label=$u$,label.dist=3,label.side=right}{v1,j4}
\fmf{fermion,label=$d$,label.side=right,label.dist=4}{j5,v1}
\fmf{fermion,label=$u$,tension=3}{i2,j2}
\fmffreeze
\fmfforce{(0,.3h)}{i2}
\fmfforce{(w,.3h)}{j2}
\fmfforce{(0,.6h)}{i3}
\fmfforce{(w,.65h)}{j4}
\fmfforce{(.55w,.6h)}{v1}
\end{fmfgraph*}
\end{fmffile}
\caption{The one diagram that contributes to the $\Delta I=3/2$ (27,1) $K\to\pi\pi$ matrix element.}
\label{fig:diagram}
\end{figure}

So far $K \to \pi\pi$ matrix elements of the $\Delta I=3/2$ (27,1) four quark operator have been calculated.  The operators that have a non-trivial $\Delta I=3/2$ part that transforms as a (27,1) are $Q_1$, $Q_2$, $Q_9$, $Q_{10}$.  It turns out by Fierz symmetry that $Q_1^{3/2}$, $Q_2^{3/2}$, $Q_9^{3/2}$, $Q_{10}^{3/2}$ are all proportional to a single operator $Q^{(27,1)(3/2)}$ \protect\cite{Sam}.
\begin{align}
Q^{(27,1)(3/2)}=3Q_{1,2}^{3/2}=2Q_{9,10}^{3/2}=&\bar{s}_a\gamma_\mu(1-\gamma^5)d_a\bar{u}_b\gamma^\mu(1-\gamma^5)u_b+\bar{s}_a\gamma_\mu(1-\gamma^5)u_a\bar{u}_b\gamma^\mu(1-\gamma^5)d_b \nonumber\\
&-\bar{s}_a\gamma_\mu(1-\gamma^5)d_a\bar{d}_b\gamma^\mu(1-\gamma^5)d_b
\end{align}
Furthermore, the Wigner-Eckhart theorem can be used to write
\begin{equation}\label{WigEck}
\langle \pi^+\pi^+ | Q'^{(27,1),3/2} | K^+\rangle=-\frac{2}{3}\langle \pi^+ \pi^0 | Q^{(27,1),3/2} | K^+\rangle
\end{equation}
where
\begin{equation}
Q'^{(27,1),3/2}=\bar{s}_a\gamma_\mu(1-\gamma^5)d_a\bar{u}_b\gamma^\mu(1-\gamma^5)d_b
\end{equation}
The matrix element on the left hand side of Equation \protect\ref{WigEck} is easier to deal with, and from this matrix element one can see that only one diagram contributes to $\Delta I=3/2$ (27,1) matrix elements.  This diagram is shown in Figure \protect\ref{fig:diagram}.

The $\Delta I=3/2$ (27,1) matrix elements have been calculated on the RBC/UKQCD $24^3 \times 64$, $L_s=16$ lattices, using 2+1 dynamical flavors and domain wall fermions, with an inverse lattice spacing of $a^{-1}=1.729(28)\text{ GeV}$.  (These lattices are described in more detail in \protect\cite{Meifeng}).  Wall sources at $t=5$ and $t=59$ are used for the kaon and two pions respectively.  Averages over periodic and antiperiodic boundary conditions are performed in order to double the effective time length and prevent contamination by `around the world' propagation.  Matrix elements are evaluated with the four quark operator at different times between $t=5$ and $t=59$.  Calculations are done with sea quark masses $m_s^{\text{sea}}=0.04$, $m_u^{\text{sea}}=m_d^{\text{sea}}=0.005,0.01$, and valence quark masses in the set $\{0.001,0.005,0.01,0.02,0.03,0.04\}$ and all possible combinations such that $m_s^{\text{val}} \ge m_l^{\text{val}}$ (where $m_l=m_u=m_d$).

\begin{figure}[htp]
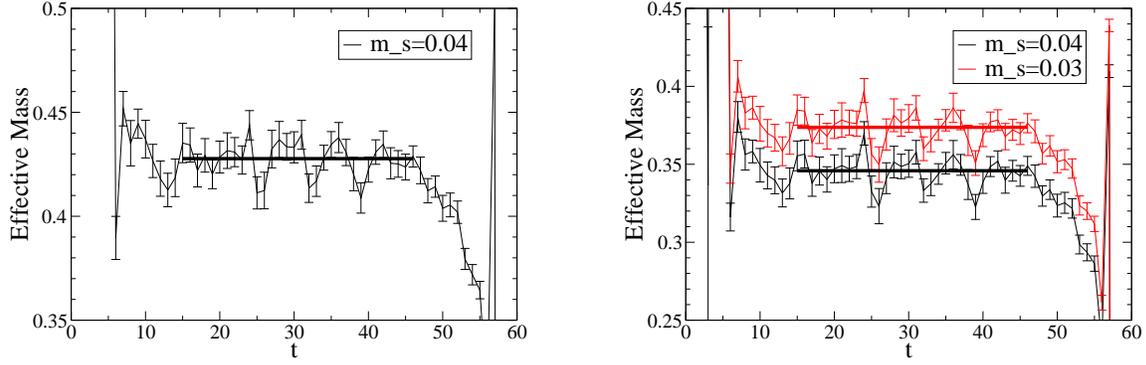

\centering
\includegraphics{effmassplot_04}
\hfill
\includegraphics{effmassplot_03}
\caption{Effective mass plots of the $K\to\pi\pi$ matrix element as a function of the time at which the four quark operator is located.  The sea quark masses are $m_l^{\text{sea}}=0.005$ and $m_s^{\text{sea}}=0.04$ in all plots here.  The valence quark masses are $m_l^{\text{val}}=0.04$ (left), $m_l^{\text{val}}=0.03$ (right), and various values of $m_s^{\text{val}}$ are shown as different curves.  A horizontal line (bold) is fitted in a range $t_{\text{min}} \le t \le t_{\text{max}}$ which is different for each curve.}\label{fig:Effmass04}
\end{figure}

\begin{figure}[htp]
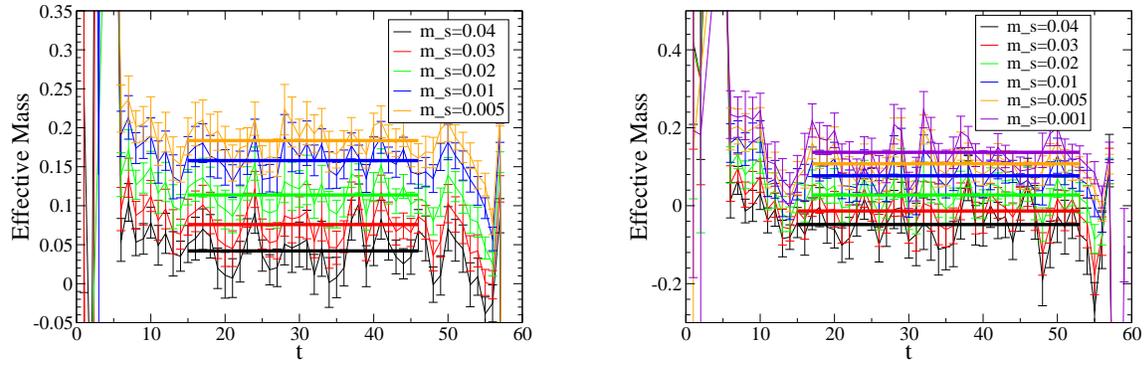

\centering
\includegraphics{effmassplot_005}
\hfill
\includegraphics{effmassplot_001}
\caption{Effective mass plots for valence quark masses $m_l^{\text{val}}=0.005$ (left), $m_l^{\text{val}}=0.001$ (right), and various values of $m_s^{\text{val}}$ shown as different curves.}\label{fig:Effmass005}
\end{figure}

Figures \protect\ref{fig:Effmass04} - \protect\ref{fig:Effmass005} show effective mass plots of the matrix element as a function of the time at which the four quark operator is located.  In these plots the sea quark masses are $m_s^{\text{sea}}=0.04$ and $m_l^{\text{sea}}=0.005$, and the valence light quark mass is held fixed while the valence strange quark mass is varied.  The plateaux are fitted in a range $t_{\text{min}} \le t \le t_{\text{max}}$ which is different for each curve, and the fits are indicated by bold lines in the plots.

The effective mass is defined as
\begin{equation}
m_{\text{eff}}(t)=\ln\left(\frac{C(t+1)}{C(t)}\right)
\end{equation}
where $C(t)$ is the value of the matrix element at time $t$.  For times far away from the sources and sinks we expect that
\begin{equation}
C(t)\approx A\exp\left[ (E_{\pi\pi}-m_K)t\right]
\end{equation}
and thus that
\begin{equation}
m_{\text{eff}}(t)\approx E_{\pi\pi}-m_K
\end{equation}
Therefore we expect the value of the plateau to be $E_{\pi\pi}-m_K$.  Results for this quantity obtained by fitting the plateaux are given in Table 2. 

\begin{table}
\label{tb:Plateaux}
\caption{Values of $E_{\pi\pi}-m_K$ obtained by fitting the plateaux of the effective mass plots.  Errors obtained from the jackknife method are quoted.  Here $m_l^{\text{sea}}=0.005$ and $m_s^{\text{sea}}=0.04$.}
\begin{center}
\begin{tabular}{|c|c|c||c|c|c||c|c|c|}
\hline
$m_s$&$m_l$&$E_{\pi\pi}-m_K$&$m_s$&$m_l$&$E_{\pi\pi}-m_K$&$m_s$&$m_l$&$E_{\pi\pi}-m_K$\\
\hline
0.04&0.04&0.4277(9)&	0.03&0.01&0.1573(20)&	0.005&0.005&0.1835(25)\\
0.04&0.03&0.3459(9)&	0.02&0.01&0.1930(20)&	0.04&0.001&-0.0483(46)\\
0.03&0.03&0.3737(10)&	0.01&0.01&0.2333(20)&	0.03&0.001&-0.0135(39)\\
0.04&0.02&0.2496(13)&	0.04&0.005&0.0420(28)&	0.02&0.001&0.0275(38)\\
0.03&0.02&0.2795(14)&	0.03&0.005&0.0758(26)&	0.01&0.001&0.0768(35)\\
0.02&0.02&0.3115(14)&	0.02&0.005&0.1138(25)&	0.005&0.001&0.1076(30)\\
0.04&0.01&0.1249(21)&	0.01&0.005&0.1579(25)&	0.001&0.001&0.1371(29)\\
\hline
\end{tabular}
\end{center}
\end{table}

\begin{figure}[htp]
\centering
\includegraphics{effmassplot_02}
\hfill
\includegraphics{effmassplot_01}
\caption{Effective mass plots for valence quark masses $m_l^{\text{val}}=0.02$ (left), $m_l^{\text{val}}=0.01$ (right), and various values of $m_s^{\text{val}}$ shown as different curves.}\label{fig:Effmass02}
\end{figure}

\section{Conclusion and Future Plans}

Calculations of $\Delta I=3/2$ (27,1) $K \to \pi\pi$ matrix elements have been done on the RBC/UKQCD 2+1 flavour dynamical $24^3 \times 64$, $L_s=16$ lattices.  To obtain physically normalized matrix elements it is necessary to use kaon, pion, and two pion correlators, the latter of which are in the process of being calculated.  Then it will be possible to move onto matrix elements of $Q_7^{3/2}$ and $Q_8^{3/2}$ which transform as (8,8).  The next step is to calculate matrix elements of the $\Delta I=1/2$ operators, however these present additional computational difficulties related to large vacuum subtractions.  

These matrix elements on the lattice must also be renormalized to their continuum values.  This can be done using the technique of non-perturbative renormalization (NPR).  Work on NPR for $K \to \pi\pi$ has been done by Shu Li.

Finally, the matrix elements must be fit to NLO $\chi PT$ expressions as soon as they become available.  This should yield the LECs necessary to calculate the physical value of the $K \to \pi\pi$ matrix element.  We would also like to calculate some matrix elements with pions that have non-zero momenta, for example the first non-vanishing value of momentum allowed by the lattice, and use this as a consistency check of the results obtained from partial quenching.  The RBC/UKQCD is also working on another large lattice of dimensions $32^3\times 64$.  Other lattice sizes could be used to perform a continuum extrapolation.

\section{Acknowledgements}

I thank Christopher Aubin, Thomas Blum, Michael Cheng, Norman Christ, Saul Cohen, Chulwoo Jung, Changhoan Kim, Min Li, Shu Li, Meifeng Lin, Robert Mawhinney, Chris Sachrajda, Amarjit Soni, and Takeshi Yamazaki for useful discussion on both physics and software.  We thank our colleagues in the RBC and UKQCD collaborations for the development and support of the QCDOC hardware and software infrastructure which was essential to this work.  In addition we acknowledge Columbia University, BNL and the U.S.\ DOE for providing the facilities on which this work was performed.  This research was supported by DOE grant DE-FG02-92ER40699.

\end{document}